\documentclass[useAMS,usenatbib]{mn2e}
\usepackage{graphicx}
\usepackage{epsfig}
\usepackage{color}
\usepackage{tabularx}
\usepackage{multirow}
\usepackage[para,online,flushleft]{threeparttable}
\usepackage[ampersand]{easylist}
\usepackage{hyperref}
\hypersetup{
    colorlinks=true,
    linkcolor=blue,
    filecolor=blue,      
    urlcolor=blue,
    citecolor=blue,
    filecolor=blue,
}
\title[X-ray and gamma-ray view of 3C 396]{{\it Suzaku} and {\it Fermi} view of the supernova remnant 3C 396}

\author[Sezer et al.]{A.~Sezer,$^{1}$\thanks{E-mail: {\color {blue}aytap.sezer@avrasya.edu.tr} (AS)}
T.~Ergin$^{2}$\thanks{{\color {blue}ergin.tulun@gmail.com} (TE)}, N.~Cesur$^{3}$\thanks{{\color {blue}nergis.raycheva@ru.nl} (NC)}, S.J.~Tanaka$^{4}$\thanks{{\color {blue}sjtanaka@phys.aoyama.ac.jp} (ST)},\\
\\
\LARGE {\rm S.~Kisaka$^{4,5}$\thanks{{\color {blue}kisaka@phys.aoyama.ac.jp} (SK)}, Y.~Ohira$^{6}$\thanks{{\color {blue}y.ohira@eps.s.u-tokyo.ac.jp} (YO)} and R.~Yamazaki$^{4,7}$\thanks{{\color {blue}ryo@phys.aoyama.ac.jp} (RY)}}\\
$^{1}$Department of Electrical-Electronics Engineering, Avrasya University, 61250, Trabzon, Turkey\\
$^{2}$TUBITAK Space Technologies Research Institute, ODTU Campus, 06800, Ankara, Turkey\\
$^{3}$Department of Astrophysics/IMAPP, Radboud University Nijmegen, 6500 GL Nijmegen, the Netherlands\\
$^{4}$Department of Physics and Mathematics, Aoyama Gakuin University, 5-10-1 Fuchinobe, Sagamihara 252-5258, Japan\\
$^{5}$Frontier Research Institute for Interdisciplinary Sciences, Tohoku University, Miyagi, 980-8578, Japan\\
$^{6}$Department of Earth and Planetary Science, The University of Tokyo, 7-3-1 Hongo, Bunkyo-ku, Tokyo 113-0033, Japan \\
$^{7}$Institute of Laser Engineering, Osaka University, 2-6 Yamadaoka, Suita, Osaka 565-0871, Japan\\
}

\begin{document}
\date{}
\pagerange{\pageref{firstpage}--\pageref{lastpage}} \pubyear{2019}
\maketitle
\label{firstpage}

%%%  ABSTRACT
\begin{abstract}
3C 396 is a composite supernova remnant (SNR), consisting of a central pulsar wind nebula (PWN) and a bright shell in the west, which is known to be interacting with molecular clouds (MCs). We present a study of X-ray emission from the shell and the PWN of the SNR 3C 396 using archival {\it Suzaku} data. The spectrum of the SNR shell is clearly thermal, without a signature of a non-thermal component. The abundances of Al and Ca from the shell are slightly enhanced, which indicates the presence of metal-enriched supernova ejecta. The PWN spectra are well described by a power-law model with a photon index of $\sim$1.97 and a thermal component with an electron temperature of $\sim$0.93 keV. The analysis of about 11-years of {\it Fermi} data revealed an 18 sigma-detection of gamma-ray emission from the location overlapping with the position of 3C 396 / 4FGL J1903.8+0531. The spectrum of 3C 396 / 4FGL J1903.8+0531 is best-fitted with a log-parabola function with parameters of $\alpha$ = 2.66 and $\beta$ = 0.16 in the energy range of 0.2$-$300 GeV. The luminosity of 3C 396 / 4FGL J1903.8+0531 was found to be $>$10$^{35}$ erg s$^{-1}$ at 6.2 kpc, which rules out the inverse Compton emission model. Possible scenarios of gamma-ray emission are hadronic emission and bremsstrahlung processes, due to the fact that the SNR is expanding into dense MCs in the western and northern regions of the SNR.
\end{abstract}

%%%  KEYWORDS
\begin{keywords}
ISM: individual objects: 3C 396 (G39.2$-$0.3), 4FGL J1903.8+0531 $-$ ISM: supernova remnants $-$ X-rays: ISM $-$ gamma-rays: ISM.
\end{keywords}

%%%  INTRODUCTION
\section{Introduction}
Composite supernova remnants (SNRs) are identified by the SNR shell and a pulsar wind nebula (PWN) (like MSH 11$-$62 and G327.1$-$1.1,  see e.g.  \citealt{Sl12, Te15}). Observations of composite SNRs provide information about the ejecta, evolution of the nebulae and structure of the surrounding circumstellar medium (CSM)/interstellar medium (ISM) (for review see \citealt{Sl17}).

The gamma-ray emission in composite SNRs may originate from the interaction between the reverse shock and the PWN (e.g. \citealt{Sl12}). The Large Area Telescope detector on board {\it Fermi} Gamma-Ray Space Telescope ({\it Fermi}-LAT) revealed gamma-rays emission from composite SNRs such as CTA 1 \citep{Ab08} and MSH 15$-$56 \citep{Te13}.    

3C 396 (also known as G39.2$-$0.3, HC24 or NRAO 593) is one of the known composite SNRs, which was first classified as a Crab-like SNR in the radio band \citep{Ca75, Ca82}. Using {\it ASCA} data, \citet{Ha99} presented the results of the X-ray study of 3C 396 and showed the unambiguous composite nature. They extracted the X-ray spectra from a circular region using a radius of about 4 arcmin that covers the entire SNR. Their analysis reveals a spectrum best described by a combination of thermal and non-thermal models. They reported that the non-thermal emission is from a central region, providing evidence for the presence of a PWN. The spectral index of the power-law component is $\Gamma$ $\sim$ 2.53 and the associated X-ray luminosity is $L_{\rm X}$ $\sim$ 2.28$\times10^{35} D_{10}^{2}$ erg s$^{-1}$ (0.2$-$4.0 keV), where $D_{10}$ is the distance to the SNR in units of 10 kpc. The thermal component arising from the interaction of the blast wave with the ISM has an electron temperature $kT_{\rm e}$ $\sim$ 0.62 keV. They found high absorption, $N_{\rm H}$ $\sim$ 4.65$\times10^{22}$ cm$^{-2}$. Their results implied a remnant age of $\sim$7000 yr in its Taylor-Sedov phase of evolution and an X-ray-emitting mass of $M_{\rm X}$ in the range of 40 and 300 $M_{\sun}$.

\citet{Ol03} analysed the 100 ks {\it Chandra} observation of 3C 396 and detected a point source at the centre of the nebula ($\rmn{RA}(J2000)=19^{\rmn{h}} 04^{\rmn{m}} 04\fs7$, $\rmn{Dec.}~(J2000)=05\degr 27\arcmin 12\arcsec$). They also detected three extensions in the east and west regions. The spectrum of the entire central nebula was characterized by a single absorbed power-law model. The best-fitting model from the 0.5$-$7.5 keV spectrum yielded an absorption column density of $N_{\rm H}$ $\sim$ 5.3$\times10^{22}$ cm$^{-2}$ and a photon index of $\Gamma$ $\sim$ 1.5. They found no thermal emission features in their analysis. From the X-ray spectral analysis of the point source, they also derived a photon index of $\Gamma$ $\sim$ 1.2. Using the Very Small Array (VSA) telescope, the possible presence of anomalous microwave emission due to spinning dust in an SNR was first suggested by \citet{Sc07}, who reported an anomalously high emission at 33 GHz. \citet{Le09} detected the [Fe\,{\sc ii}] 1.64 $\mu$m and H$_2$ 2.12 $\mu$m filaments with the near-infrared (IR) [Fe\,{\sc ii}] and H$_2$ Êline imaging and spectroscopic data of the SNR. These filaments are close to each other in the western rim of 3C 396. 

\citet{Su11} investigated the molecular environment and re-analysed {\it Chandra} ACIS observation of 3C 396. Using CO millimeter observations, they found that the western boundary of the remnant is perfectly confined by the western molecular wall at the local standard of rest (LSR) velocity of $V_{\rm LSR}$ $\sim$ $84$ km s$^{-1}$ and the multiwavelength properties are consistent with the presence of the $84$ km s$^{-1}$ molecular clouds (MCs). Along its western edge, the SNR has a broadened molecular emission, and also shows presence of OH maser at the same velocity. \citet{Su11} also extracted {\it Chandra} X-ray spectra from five X-ray bright regions excluding the PWN and the point source.  They used an absorbed non-equilibrium ionization (NEI) model, found Si and S for all regions, but Ar and Ca lines are prominent only in the northern and southern regions, where the absorption column density, $N_{\rm H}$, in the range of $\sim$ $(4.0-6.5)\times10^{22}$ cm$^{-2}$ and the electron temperature, $kT_{\rm e}$, in the range of 0.66$-$1.33 keV. They estimated a progenitor mass of 13$-$15 $M_{\sun}$ and the SNR age to be $\sim$3 kyr. In the study of \citet{Ki16}, the presence of the velocity-broadened $^{12}$CO $J=2-1$  emission centred around $+69$ km s$^{-1}$ to the north of 3C 396 was detected and it was shown to be extending to the west, where the broadened emission is centred around $+72$ km s$^{-1}$. Recently, \citet{No18} searched for the iron K-shell line in some SNRs using the {\it Suzaku} data. They found that the flux ratio of Fe\,{\sc i} K$\alpha$/Fe\,{\sc xxv} He$\alpha$ in 3C 396 is consistent with the Galactic ridge X-ray emission (GRXE) within 1$\sigma$ errors.

From the H\,{\sc i} absorption, \citet{Ca75} derived a lower limit for the distance of $\sim$7.7 kpc to 3C 396. The X-ray H\,{\sc i} column density and CO associations implied a distance of 6.2$-$8 kpc \citep{Ol03, He09, Su11}. \citet{Su11} determined a distance of $\sim$6.2 kpc, with the LSR velocity of $\sim$84 km s$^{-1}$. We assume the distance to 3C 396 to be 6.2 kpc throughout the paper\footnote{Very recently, \citet{RaLe18} updated the distance estimate as 8.5 kpc corresponding to the velocity of 69.4 km s$^{-1}$.}.

3C 396 was searched in TeV and GeV gamma-ray bands by HEGRA and H.E.S.S. and by {\it Fermi}-LAT, respectively. In the 1st {\it Fermi}-LAT Supernova Remnant Catalog \citep{Ac16}, the upper limits on the flux of this SNR were reported on Table 3, i.e. among the not-detected SNRs. A preliminary analysis of {\it Fermi}-LAT data by \citet{Er16} revealed an excess of GeV gamma rays from 3C 396. Recently, the Fourth {\it Fermi}-LAT sources (4FGL; \citealp{Fe19}) catalog contained a new gamma-ray source, 4FGL J1903.8+0531, which was detected at R.A.(J2000) = 19$^h$ 03$^m$ 54$^s\!$.0 and decl.(J2000) = $+$05$^{\circ}$ 31$'$ 17$''\!$.0, about 0$^{\circ}\!\!$.08 away from 3C 396, with a significance of 10$\sigma$. HEGRA and H.E.S.S. did not detect 3C 396 in TeV gamma-ray energies \citep{Bo11, Ah01, Ah05, Ab18}.

In this work, we analyse the high spectral resolution data from {\it Suzaku} to identify the nature and the spectral properties of the X-ray emission from 3C 396.  We also analyse {\it Fermi}-LAT data to examine a gamma-ray emission in the GeV energy range. The paper is organized as follows. In Section \ref{obs}, we summarize X-ray and gamma-ray observations and data reduction. In Section \ref{analysis}, we describe how we estimated background components, explain our spectral analysis procedure and give results of the X-ray and gamma-ray analysis. In Section \ref{discussion}, we discuss the nature of X-ray and gamma-ray emission and the spectral properties of the SNR shell and the interior of the remnant. Finally, we draw our conclusions in Section \ref{conclusions}. 

%%%  OBSERVATION AND DATA REDUCTION
\section{Observations and Data Reduction}
\label{obs}

%% X-ray Data
\subsection{X-ray data}
3C 396 was observed with {\it Suzaku} X-ray Imaging Spectrometer (XIS; \citealt{Ko07}) on 2014 April 26 (ObsID: 509038010; PI: T. Pannuti). The net exposure of the cleaned event data was $\sim$72 ks. We used the data obtained with front-side illuminated (FI) CCD chips (XIS0 and XIS3) and back-side illuminated (BI) chip (XIS1). 

The X-ray data reduction and analysis were performed with {\sc heasoft}\footnote{\url{https://heasarc.nasa.gov/lheasoft}} version 6.20 and {\sc xspec} version 12.9.1 \citep{Ar96} with atomic data base ({\sc atomdb}) version 3.0.9\footnote{\url{http://www.atomdb.org}} \citep{Sm01,Fo12}. We analysed cleaned event data that had already been preprocessed with the {\it Suzaku} team. To analyse the XIS data, we generated redistribution matrix files with ftool {\sc xisrmfgen} and ancillary response files with {\sc xissimarfgen} \citep{Is07}. The spectra were grouped to include at least 25 counts in each bin.
%FIGURE 1
\begin{figure*}
\centering \vspace*{1pt}
\includegraphics[width=0.52\textwidth]{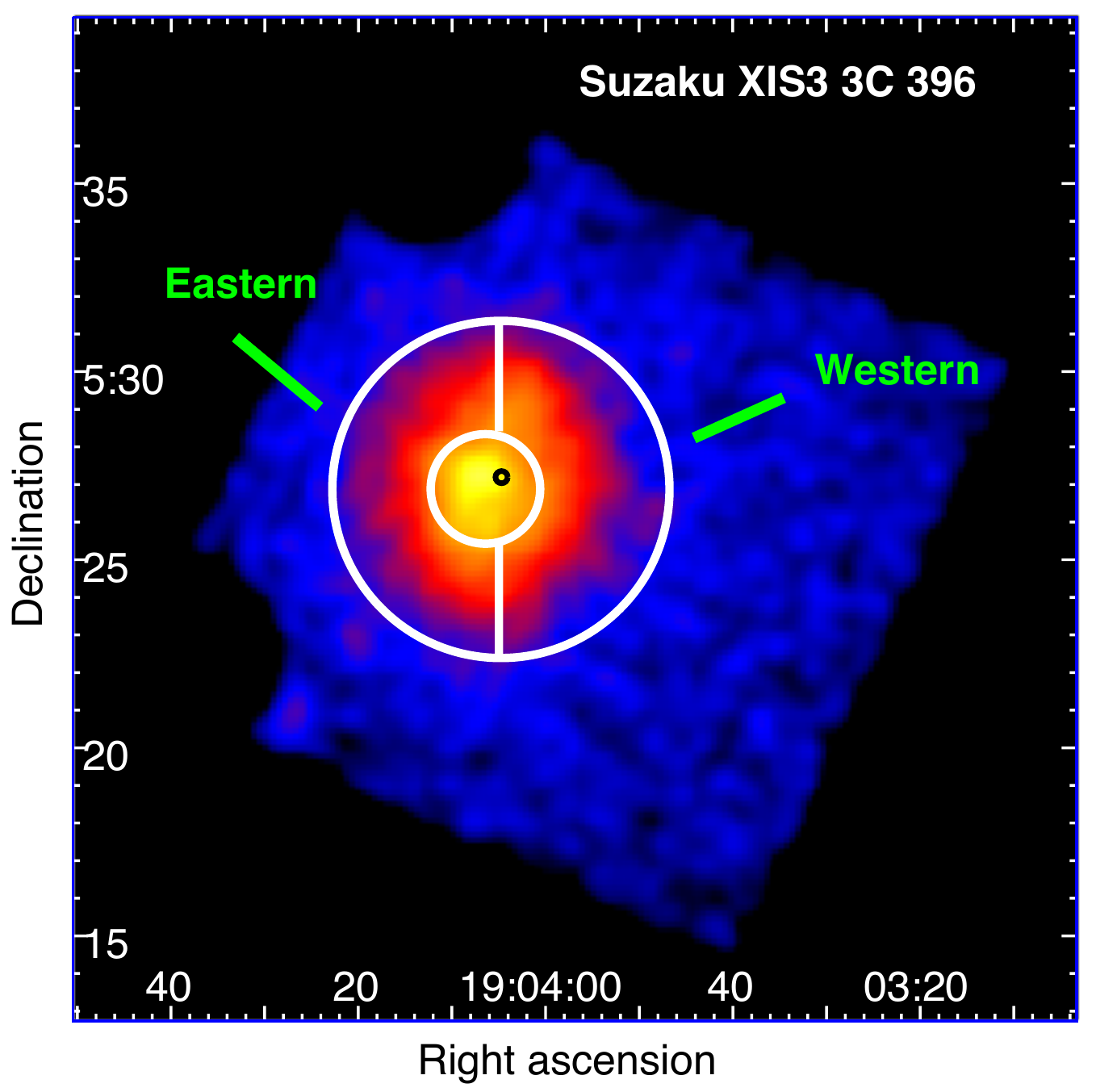}
\caption{{\it Suzaku} XIS3 image of 3C 396 in the 0.3$-$10.0 keV energy band. The $^{55}$Fe calibration source regions at two CCD corners of the FoV ($\sim$17.8$\times$17.8 arcmin$^{2}$) were excluded. The circles with radii of 1.5 and 4.5 arcmin indicate the extracted regions for the spectral analysis. The lines divide the annulus into two regions for further analysis. The smallest circle marks the pulsar's location.}
\label{Fig1}
\end{figure*}

%% Gamma-ray Data Reduction
\subsection{Gamma-ray data reduction}
The gamma-ray observations were taken from 2008-08-04 to 2019-10-07.  In this analysis, we made use of the analysis packages \texttt{fermitools}\footnote{\url{http://fermi.gsfc.nasa.gov/ssc/data/analysis/software}} version \texttt{1.0.1} and  \texttt{fermipy}\footnote{\url{http://fermipy.readthedocs.io/en/latest/index.html}} version \texttt{0.17.4}. Using \texttt{gtselect} of \texttt{fermitools} we selected {\it Fermi}-LAT Pass 8 `source' class and `front$+$back' type events coming from zenith angles smaller than 90$^{\circ}$ and from a circular region of interest (ROI) with a radius of 20$^{\circ}$ centered at the SNR's radio location\footnote{SNR's radio location at  R.A.(J2000) = 285$^{\circ}\!\!$.975 and decl.(J2000) = 5$^{\circ}\!\!$.521}. The {\it Fermi}-LAT instrument response function version \emph{P8R3$_{-}$SOURCE$_{-}\!\!$V2} was used. For mapping the morphology and searching for new sources within the analysis region, events having energies in the range of 1$-$300 GeV were selected. To deduce the spectral parameters of the investigated sources, events with energies between 200 MeV and 300 GeV were chosen.

%%% ANALYSIS AND RESULTS
\section{Analysis and Results}
\label{analysis}
% FIGURE 2
\begin{figure*}
\centering \vspace*{1pt}
\includegraphics[width=0.49\textwidth]{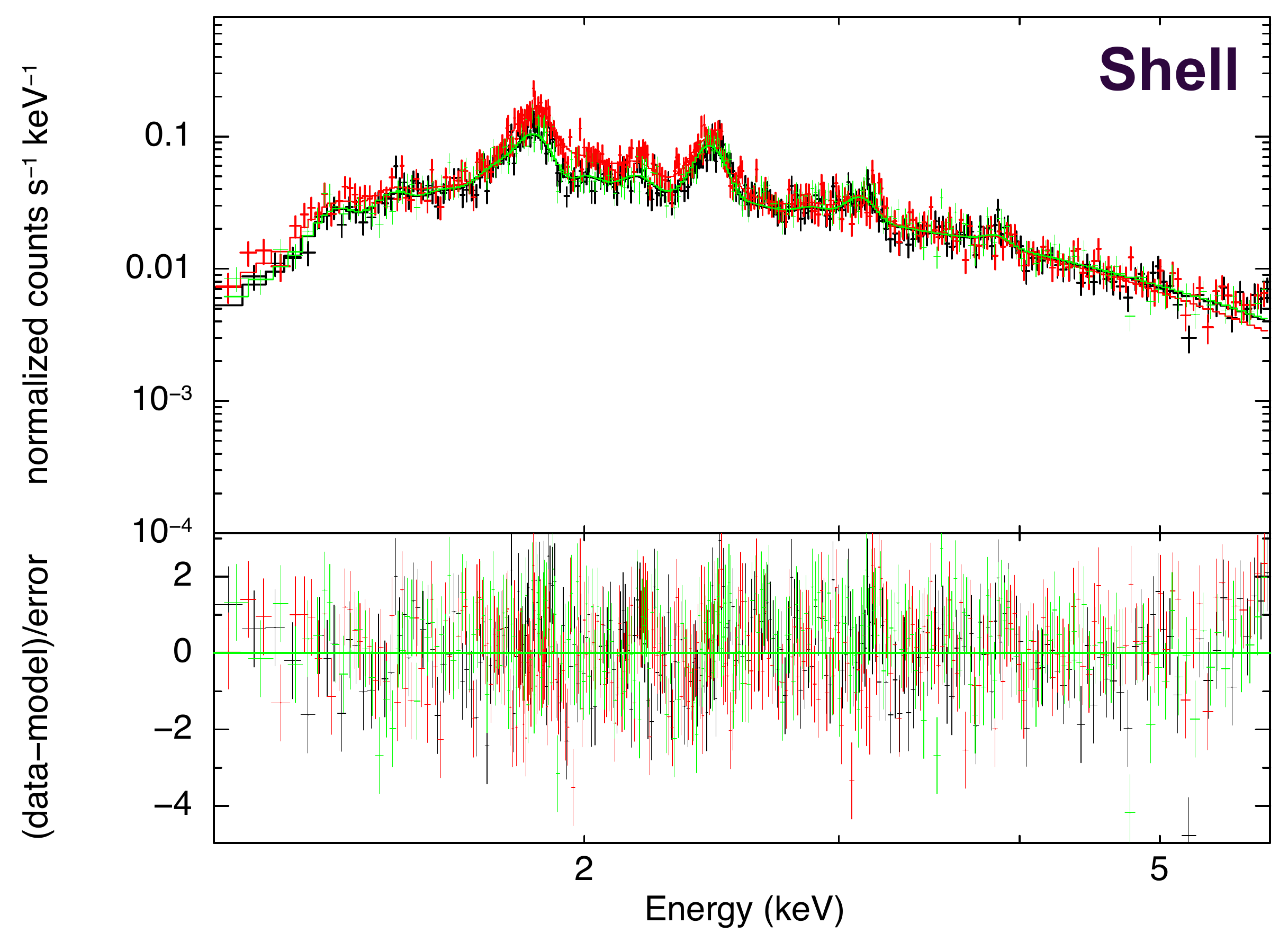}
\includegraphics[width=0.49\textwidth]{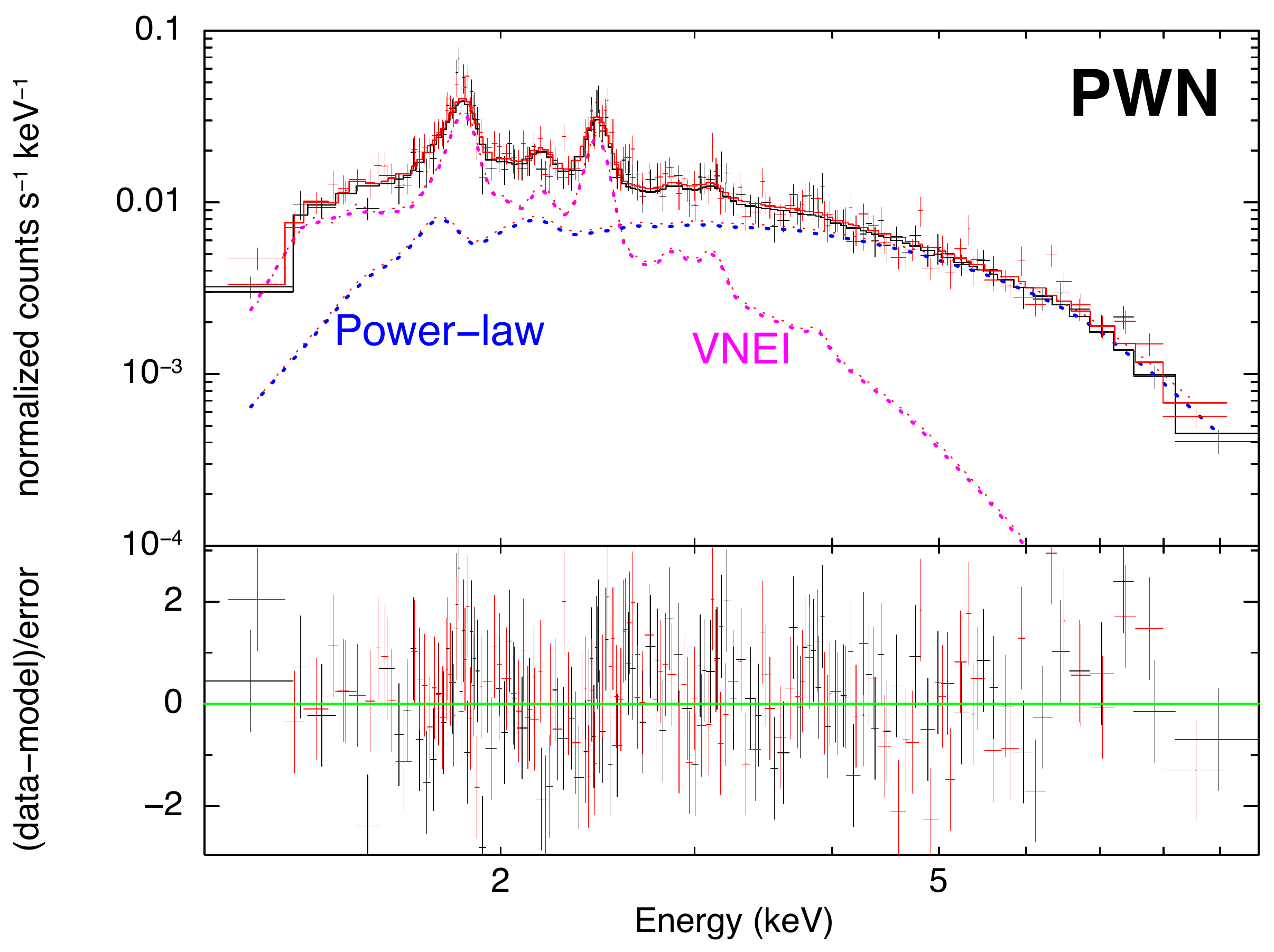}
\caption{{\it Suzaku} XIS (XIS0, 1 and 3) spectra of the SNR shell in the 1.0$-$6.0 keV and the XIS FI (XIS0 and 3) spectra of the PWN in the 1.0$-$10.0 keV energy band, fitted with the model described in Section 3.2. The spectra are overlaid with the best-fitting model. The fit residuals are shown in the lower part of each panel.}
\label{Fig2}
\end{figure*}

%% X-ray Spectral Analysis
\subsection{Analysis of {\it Suzaku} data}

% XIS Image
\subsubsection{XIS image} 
Fig. \ref{Fig1} displays an XIS3 image of 3C 396 in the 0.3$-$10.0 keV energy range, where the spectral analysis regions are: (i) The outermost circular region defining the SNR shell extraction region, excluding the PWN and the point source region; (ii) The eastern and western shells of the SNR; (iii) The $3.0$ arcmin-diameter circle defining the PWN region, excluding the innermost circle contaminated by emission from the point source.

% TABLE 1
\begin{table*}
\begin{minipage}{170mm}
\begin{center}
\caption{{\it Suzaku} best-fitting spectral parameters for the SNR shell and the PWN.}
\begin{threeparttable}
\renewcommand{\arraystretch}{1.5}
\begin{tabular}{@{}p{2.9cm}p{3.9cm}p{2.9cm}p{2.9cm}@{}}
\hline\hline
Component & Parameters                              & Shell                           & PWN                              \\
\hline
TBABS  &$N_{\rm H}$ ($10^{22}$ cm$^{-2})$     	& $5.83_{-0.24}^{+0.27}$          &  $5.22_{-0.17}^{+0.24}$              \\

VVNEI  &  $kT_{\rm e}$ (keV)                    & $1.12_{-0.03}^{+0.02}$          &  $0.93_{-0.09}^{+0.08}$              \\

&  Mg (solar)                                      	& $1.2_{-0.2}^{+0.2}$             &            (1)                       \\

&  Al (solar)                                      	& $1.9_{-0.3}^{+0.4}$             &            (1)                       \\

&  Si (solar)                                     	& $1.3_{-0.2}^{+0.1}$             &  $1.5_{-0.2}^{+0.2}$            	\\

&  S (solar)                                     	&  $1.1_{-0.2}^{+0.2}$            &  $1.4_{-0.3}^{+0.1}$     		\\

&  Ca (solar)                                     	& $1.7_{-0.2}^{+0.4}$             &           (1)                	\\

&  $\tau$ ($10^{11}$ cm$^{-3}$ s)   	 	& $2.3_{-0.5}^{+0.4}$             &  $3.6_{-0.4}^{+0.4}$    		\\

& Norm$\dagger$  ($10^{-3}$ cm$^{-5}$)       		& $38.12_{-2.49}^{+4.08}$         & $4.79_{-0.32}^{+0.65}$   		\\

Power-law&  $\Gamma$                                	&              $-$                    & $1.97_{-0.38}^{+0.23}$    		\\

&  Norm$\ddagger$ ($10^{-5}$) 				&         	 $-$	          & $1.64_{-0.45}^{+0.28}$ 		\\

& Reduced $\chi^{2}$ (dof)                      & 1.12 (828)                      &  1.03 (379)                         \\
 \hline
\end{tabular}
\begin{tablenotes}
\item {\bf Notes.} Errors are within a 90 per cent confidence level. Abundances are given relative to the solar values of \citet{Wi00}.
(1) indicates that the elemental abundance was fixed at solar. \\
\item $\dagger$ The normalization of the VVNEI, norm=$10^{-14}$ $\int n_{\rm e} n_{\rm H} dV$/($4\pi d^{2}$), where $d$ is the distance to the source (in cm), $n_{\rm e}$, $n_{\rm H}$ are the electron and hydrogen densities (in units of cm$^{-3}$), respectively, and $V$ is the emitting volume (in units of cm$^{3}$). \\
\item $\ddagger$ The normalization of the power-law is in units of photons cm$^{-2}$ s$^{-1}$ keV$^{-1}$ at 1 keV. 
\end{tablenotes}
\end{threeparttable}
\end{center}
\end{minipage}
\end{table*}

%% Background Estimation
\subsection{Background estimation}
For 3C 396, \citet{Ol03} and \citet{Su11} extracted the background spectra by selecting regions from the field of view (FoV) of the {\it Chandra} observation in their analysis. We extracted the background spectra from the entire source-free region of the same chip area, excluding the calibration regions and the SNR region. The instrumental (non-X-ray) background (NXB) spectra of XIS were generated using {\sc xisnxbgen} \citep{Ta08}. The CXB and GRXE are considered as astrophysical X-ray background components. The background spectrum was fitted with the following model:

% FIGURE 3
\begin{figure*}
\centering \vspace*{1pt}
\includegraphics[width=0.99\textwidth]{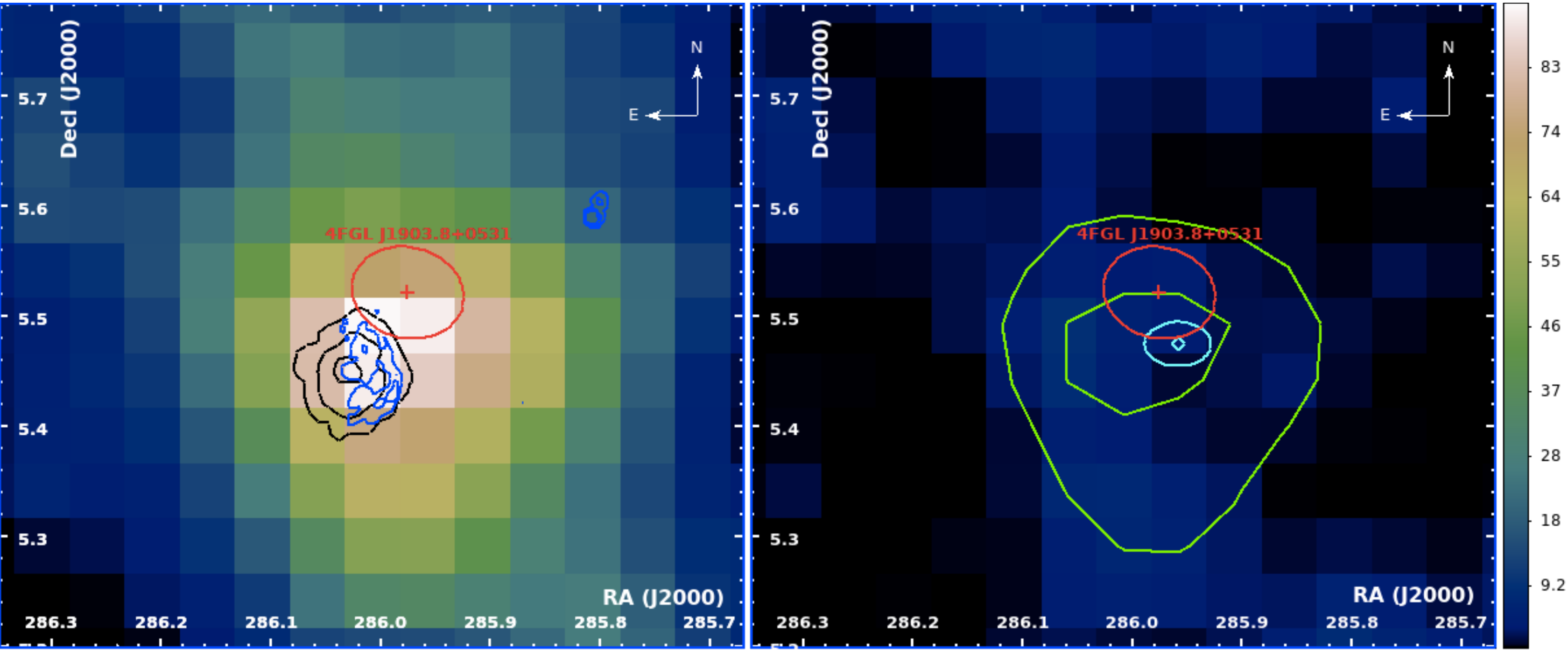}
\caption{Distribution of GeV gamma-ray emission from the neighborhood of 3C 396. Both TS maps have a bin size of 0$^{\circ}\!\!$.05$\times~$0$^{\circ}\!\!$.05 and their color scales are adjusted to the TS range of [1,92]. The red plus markers and red ellipses show the point source locations and position errors of the 4FGL catalog sources, respectively. {\bf Left Panel:} TS map produced without the inclusion of 4FGL J1903.8+0531 into the gamma-ray background model. The black dashed contours represent the X-ray flux values from {\it Suzaku}, which are 9, 12, 19, and 25 photons cm$^{-2}$ s$^{-1}$. The blue solid contours are radio data, 20 cm VLA \citep{Be87}, at 2, 5, and 9 mJy per beam. {\bf Right Panel:} The TS map produced including 4FGL J1903.8+0531, as a point source at the best-fit position, into the gamma-ray background model. The best-fit position is shown with a cyan diamond and the statistical error of this position is given with a cyan ellipse. Green solid contours represent gamma-ray significance values of 7$\sigma$ and 9$\sigma$.}

\label{Fig3}
\end{figure*}

\begin{equation}
{\rm Abs1} \times ({\rm power}\mbox{-}{\rm law}) + {\rm Abs2} \times ({\rm apec} + {\rm apec}),
\end{equation}
where Abs1 and Abs2 represent the ISM absorption for CXB and GRXE, respectively. The apec is a collisional ionisation equilibrium (CIE) plasma model in the \textsc{xspec}. The CXB component parameters are fixed at those in \citet{Ku02}, where power-law fit with a photon index of ${\Gamma}=1.41$ and a surface brightness of 5.41$\times$10$^{-15}$ erg s$^{-1}$ cm$^{-2}$ arcmin$^{-2}$ in the 2.0$-$10.0 keV energy band were adopted. The absorption column density is assumed to be $N_{\rm H (CXB)}$=2 $\times$ $N_{\rm H (GRXE)}$ \citep{Uc13}. The GRXE component is represented with two CIE models (apec+apec) as a low-temperature plasma ($kT_{\rm e}$ $\sim$ 1 keV) and a high-temperature plasma ($kT_{\rm e}$ $\sim$ 7 keV) \citep{Uc13}. We generated a background spectrum file using the \textsc{fakeit} command in \textsc{xspec} and used it for each source spectrum as background. 

% The SNR Shell
\subsubsection{The SNR Shell}
To examine the X-ray spectral properties of the SNR shell, we extracted spectra of the entire SNR from a circular region with a radius of 4.5 arcmin, shown by the outermost circle in Fig. \ref{Fig1}. Emissions from the PWN and the point source were eliminated from the spectra.

After subtracting the background, the SNR has no significant X-ray emission above $\sim$6 keV, which is consistent with the result of \citet{No18}. We employed an absorbed single-component variable-abundance NEI plasma model (VVNEI in {\sc xspec}), where the free parameters were the normalization, the absorption column density ($N_{\rm H}$), the electron temperature ($kT_{\rm e}$), the ionization parameter ($\tau$=$n_{\rm e}t$) and the abundances of Mg, Al, Si, S and Ca. Other metal abundances were fixed to the solar values \citep{Wi00}. We found an electron temperature of $kT_{\rm e}$=$1.12_{-0.03}^{+0.02}$ keV, an ionization time-scale of $n_{\rm e}t$=$2.3_{-0.5}^{+0.4}\times10^{11}$ cm$^{-3}$ s, an absorption column density of $N_{\rm H}$=$5.83_{-0.24}^{+0.27}$ $\times10^{22}$ cm$^{-2}$, with evidence for slightly enhanced abundances of Al and Ca. The {\it Suzaku} XIS spectra of the SNR shell in the 1.0$-$6.0 keV are illustrated in Fig. \ref{Fig2} (left panel). The best-fitting parameters are given in Table 1. 

To better understand the emission characteristics of the SNR shell surrounding the PWN, we also extracted XIS spectra from two selected regions (as shown in Fig. \ref{Fig1}), which correspond to the X-ray, IR and radio - bright western and faint eastern shells. For the western shell, we obtained an absorption column density of $N_{\rm H}$=$6.03_{-0.34}^{+0.47}$$\times10^{22}$ cm$^{-2}$, which is higher than that for the eastern region.

% The PWN
\subsubsection{The PWN}
We extracted the PWN spectra from an X-ray bright region with a 1.5 arcmin radius circle excluding the point source ($r$=0.16 arcmin). We fitted the spectra of the PWN with an absorbed VVNEI and power-law model. The spectra are well described by a power-law model with $N_{\rm H}$=$5.22_{-0.17}^{+0.24}$ $\times10^{22}$ cm$^{-2}$ and $\Gamma$=$1.97_{-0.38}^{+0.23}$ and a thermal component with an electron temperature of $0.93_{-0.09}^{+0.08}$ keV. The best-fitting spectral parameters are summarized in Table 1 and the XIS spectra of the PWN in the 1.0$-$10.0 keV energy band are given in Fig. \ref{Fig2} (right panel).

%% Gamma-ray Analysis and Results
\subsection{Gamma-ray analysis and results}
The background model of the analysis region consists of diffuse background sources, diffuse Galactic emission (GE) and the isotropic component (ISO), and all the extended and point sources from 4FGL Catalog \citep{Fe19} located within a 10$^{\circ}$ $\times$ 10$^{\circ}$ region centred on the ROI centre. We freed all parameters of GE (\emph{gll$_{-}$iem$_{-}$v7.fits}) and ISO (\emph{iso$_{-}$P8R3$_{-}$SOURCE$_{-}\!\!$V2$_{-}\!$v1.txt}). The normalization parameters of all sources within 3$^{\circ}$ are set free. In addition, we freed all sources that have test statistic (TS\footnote{TS = -2ln($L_{\rm max,0}$/$L_{\rm max,1}$), where $L_{\rm max,0}$ is the maximum likelihood value for a model without an additional source and $L_{\rm max,1}$ is the maximum likelihood value for a model with the additional source at a given location.}) values greater than 400 and fixed all sources with TS values smaller than 400.

% Position and Source Morphology
\subsubsection{Position and Source Morphology}
In this analysis, we used the gamma rays in the energy range of 1$-$300 GeV. We removed 4FGL J1903.8+0531 from the gamma-ray background model and the resulting distribution of the gamma-ray emission presents a point-like source morphology as shown on the TS map given in Fig. \ref{Fig3} left panel. The best-fit position of the excess gamma-ray emission was found to be R.A.(J2000) = 285$^{\circ}\!\!$.957 $\pm$ 0$^{\circ}\!\!$.027 and decl.(J2000) = 5$^{\circ}\!\!$.475 $\pm$ 0$^{\circ}\!\!$.021 using the \texttt{localize} method of the \texttt{fermipy} analysis package. This position is 0$^{\circ}\!\!$.08 away from the radio location of 3C 396 and only 0$^{\circ}\!\!$.05 away from 4FGL J1903.8+0531, where the error ellipse of the best-fit position and the positional error of 4FGL J1903.8+0531 are overlapping. Therefore, herewith we name the excess of gamma rays found in this analysis as 4FGL J1903.8+0531 and/or 3C 396. If we insert 4FGL J1903.8+0531 back into the gamma-ray background model using the best-fit position found in this analysis, we obtain the TS map given in Fig. \ref{Fig3} right panel.

% Gamma-ray Spectrum
\subsubsection{Gamma-ray Spectrum}
The spectral measurement was performed in the 200 MeV$-$300 GeV energy range, where the spectrum of 4FGL J1903.8+0531 was fit to the Log-parabola (LP) spectral model below:
 \begin{equation}
\mbox{dN/dE} =  \mbox{N}_0 ~(\mbox{E}/\mbox{E}_b)^{{\rm-(\alpha + \beta \log(E/E_{b}))}}, 
\end{equation}
where E$_b$ is a scale parameter. ${\rm \beta}$ and ${\rm \alpha}$ are spectral indices of the LP spectral model. N$_0$ is the normalization parameter. 

\begin{figure*}
\centering 
\includegraphics[width=0.7\textwidth]{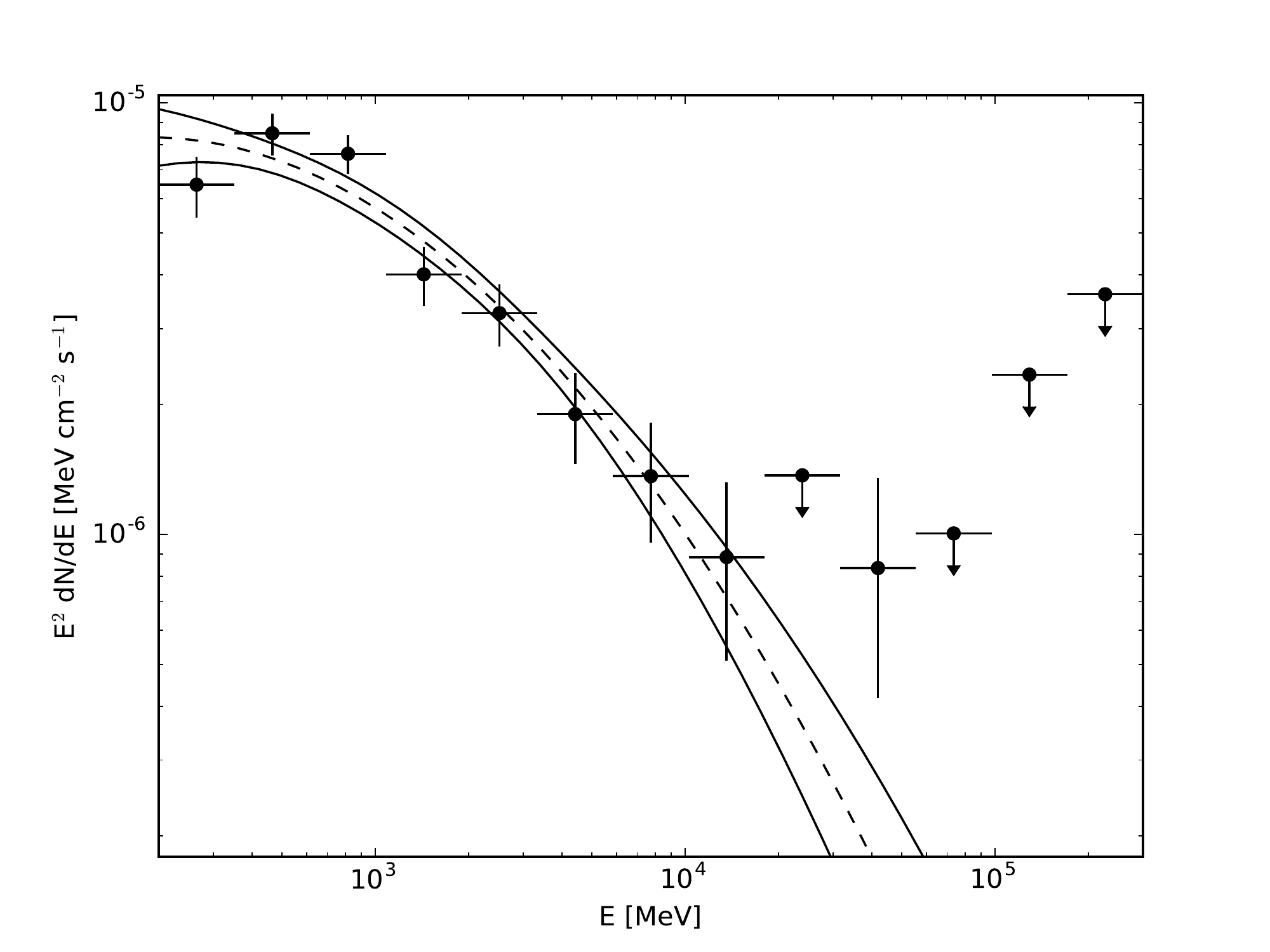}
\caption{Gamma-ray energy distribution of 4FGL J1903.8+0531 / 3C 396 which is fit with a LP-type spectrum between 200 MeV and 300 GeV. The upper limits at 95 per cent confidence level on the energy flux. The area inside the two solid lines represent the error band and its center is shown by the dashed line.}
\label{Fig4}
\end{figure*}

The TS value of 4FGL J1903.8+0531 in this energy range was found to be higher (TS = 338) than what was reported in the 4FGL catalog \citep{Fe19}. The gamma-ray spectral data points and their statistical error bars are shown in Fig. \ref{Fig4}. The fitted LP-type spectral model is drawn on top of the data points. The LP-spectral indices were found to be ${\rm \alpha}$ = 2.66 $\pm$ 0.09 and ${\rm \beta}$ = 0.16 $\pm$ 0.05. In the energy range of 0.2$-$300 GeV, the total flux and energy flux values were found to be (3.19 $\pm$ 0.26)$\times$10$^{-8}$ cm$^{-2}$ s$^{-1}$ and (1.89 $\pm$ 0.12)$\times$10$^{-5}$ MeV cm$^{-2}$ s$^{-1}$, respectively. Comparing these results with the spectral parameters derived for 4FGL J1903.8+0531 in the 4FGL catalog are ${\rm \alpha}$ = 2.72 $\pm$ 0.16, ${\rm \beta}$ = 0.26 $\pm$ 0.08, and energy flux = (1.94 $\pm$ 0.35)$\times$10$^{-5}$ MeV cm$^{-2}$ s$^{-1}$, we conclude that the parameters of both analyses are compatible with each other.  

%%% DISCUSSION
\section{Discussion}
\label{discussion}
We present a study of emission from the SNR 3C 396 using the {\it Suzaku} and {\it Fermi}$-$LAT data. In this section, we briefly discuss the implication of our results.

\subsection{Thermal X-ray emission}
The {\it Suzaku} X-ray spectra of the SNR shell are well described by a thermal plasma model with an electron temperature of $kT_{\rm e}$ $\sim$ 1.1 keV and an ionization time-scale of $\tau$ $\sim$ $2.3\times10^{11}$ cm$^{-3}$ s, indicating that the plasma is still ionizing. We clearly detected K-shell line of Al from the shell. The overabundant Al was found in several core-collapse SNRs (e.g. G350.1$-$0.3: \citealt{Ya14}; Kes 17: \citealt{Wa16}). Our abundance patterns are not consistent with the ones reported in \citet{Su11}. This may be due to the fact that we extracted spectra from a larger region in comparison to \citet{Su11}. The reason for this difference may be, because more recent atomic data is implemented in this work.

% FIGURE 4

The absorption column density, $N_{\rm H}$ $\sim$ 5.8$\times10^{22}$ cm$^{-2}$, inferred from the TBABS model is slightly larger than that obtained by \citet{Ha99, Ol03, Su11} from their WABS model \citep{Mo83} fit to {\it ASCA} and {\it Chandra} data, because we used the lower metallicity of the solar abundance set.

We also examined the X-ray spectral properties of the PWN with the XIS data. The spectrum can be described by an NEI model with an electron temperature $\sim$0.93 keV, ionization time-scale $\sim$ $3.6\times10^{11}$ cm$^{-3}$ s and power-law model with a photon index $\sim$1.97. The thermal emission might come from the front and back of the SNR shell as previously explained by \citet{Ol03} or its origin might be physically related to the origin of the non-thermal emission. However, the nature of the thermal emission needs to be further studied.

% FIGURE 5
\begin{figure*}
\centering 
\includegraphics[width=0.7\textwidth]{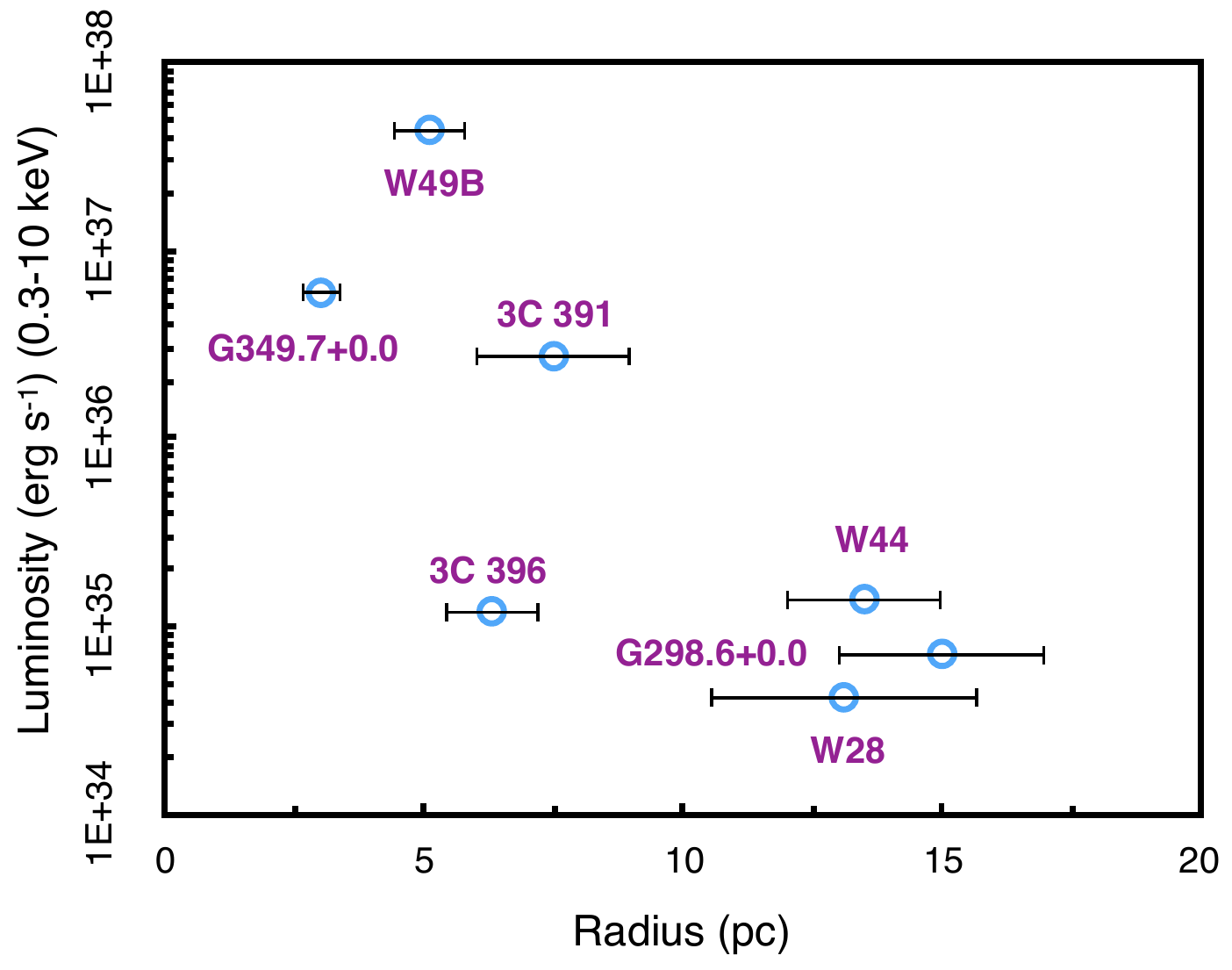}
\caption{Radius versus absorption-corrected X-ray luminosity for MM SNRs with GeV gamma-ray emission from \citet{Ba16} and 3C 396 from our work using XIS data.}
\label{Fig5}
\end{figure*}

In the following, we estimate the properties of the remnant using the normalization obtained with the VVNEI model (see Table 1) in the analysis of the XIS data, considering the distance of 6.2 kpc. We estimate the volume for the SNR shell to be  $V\sim 1.58\times10^{58}fd_{6.2}^{3}$ ${\rm cm^{3}}$, where $f$ is the volume filling factor ($0<f<1$) and $d_{6.2}$ is the distance in units of 6.2 kpc. Assuming $n_{\rm e}=1.2n_{\rm H}$, we found an ambient gas density of $n_{\rm e}$ $\sim$ 1.15$f^{-1/2}d_{6.2}^{-1/2}$ ${\rm cm}^{-3}$ and age of $\sim 6340f^{1/2}d_{6.2}^{1/2}$ yr. Our estimated SNR age is in agreement with previous age estimates (i.e. $\sim$7000 yr: \citealt{Ha99} and $\sim$3000 yr: \citealt{Su11}). Using $M_{\rm X}$=1.4$m_{\rm H}n_{\rm e}V$, we calculate the total X-ray-emitting mass, $M_{\rm X}$ $\sim 21.4 f^{1/2}d_{6.2}^{5/2}~{M\sun}$, which is lower than that derived by \citet{Su11} in their {\it Chandra} analysis ($M_{\rm X}$ $\sim 70 f^{1/2}d_{6.2}^{5/2}~{M\sun}$).

As a next step, we derive the electron density from the eastern and western regions of 3C 396 using XIS data. The obtained density is higher in the western ($\sim$2.43 ${\rm cm}^{-3}$) than in the eastern region ($\sim$0.93 ${\rm cm}^{-3}$).

In this analysis, the derived electron densities of the thermal X-ray plasma were found to be about 1$-$2 cm$^{-3}$, which is consistent with the result found by \citet{Su11} ($\sim$1 cm$^{-3}$).

We discuss the SNR morphology. For this, we checked the morphology of the radio synchrotron emission. The radio emission is much brighter in the west than in the east, with only a faint tail extending from the prominent western shell toward east (e.g. \citealt{Su11, Cr16}). 3C 396 is a young SNR that contains a PWN at its centre, which is related to non-thermal X-ray emission. This characteristics differentiates this SNR from MM SNRs, which show centrally enhanced thermal X-ray emission (e.g. \citealt{Rh98}). Additionally, unlike in most of the MM SNRs, there are significant electron temperature variations within 3C 396.

We compare the thermal X-ray luminosity with those of some MM SNRs similar to \citet{Ba16}. They plotted the radius versus thermal X-ray luminosity (0.3$-$10.0 keV) of MM SNRs with associations to GeV gamma rays and concluded that the value of the X-ray luminosity become smaller when SNRs evolve or when their radii become larger. Assuming a distance of 6.2 kpc, we derived an unabsorbed thermal X-ray luminosity in the 0.3$-$10.0 keV energy band of $\sim$ $1.2\times10^{35}$ erg s$^{-1}$ for 3C 396 using XIS data. In Fig. \ref{Fig5}, we give a plot as Fig3(b) of \citet{Ba16}, including the SNR 3C 396. Compared with other remnants, namely 3C 391, the thermal X-ray luminosity of 3C 396 is rather low despite its small radius. This could be due to absorption of soft X-ray emission from a lower temperature plasma, which is generated by shock-cloud interactions and cannot be identified in our analysis. Also, in Fig. \ref{Fig5}, the other SNRs are MM, some of which show RPs (e.g. \citealt{Oz09}).

\subsection{Gamma-ray emission}

In the MeV to GeV energy range, gamma rays are produced by hadronic and/or leptonic processes. In the hadronic process, gamma rays are produced through the $\pi^0$ decay channel originating from interactions of accelerated protons with molecular material surrounding the SNR. The leptonic gamma-ray emission mechanism, on the other hand, requires accelerated electrons and ambient photons (e.g from the cosmic microwave background (CMB)) to produce gamma rays through inverse Compton (IC) scattering and non-thermal bremsstrahlung processes. The IC scattering process is accompanied by the synchrotron emission observed in radio and X-ray wavelengths.  

As seen on the TS map given in Fig. \ref{Fig3} left panel, the distribution of gamma rays in the energy range of 1$-$300 GeV overlaps with the western and northern parts of 3C 396. The best-fit position is pointing in a direction that coincides with dense MCs reported by \citet{Su11} and \citet{Ki16}. This implies that the origin of GeV gamma rays may be due to the interaction of accelerated protons emanating from the SNR’s western or northern shell with the material inside the MCs. 

From the point source flux observed by {\it Chandra}, \citet{Ol03} discussed that the pulsar powering the PWN is a typical young pulsar of $L_{\rm sd} \sim 7.2 \times 10^{36}$ erg s$^{-1}$. Combined this with the fluxes obtained from {\it Suzaku} ($2.09^{+0.08}_{-0.11} \times 10^{-12}$ erg cm$^{-2}$ s$^{-1}$ in 1$-$10 keV and {\it Fermi} ($1.89 \times 10^{-5}$ MeV cm$^{-2}$ s$^{-1}$ in 0.2 -- 300 GeV), we obtain the X-ray and GeV $\gamma$-ray efficiencies of the PWN to be $\eta_{\rm PWN,X} \approx 10^{-2.9}$ and $\eta_{\rm PWN, GeV} \approx 10^{-1.8}$. The flux upper limit in TeV obtained from Fig. A.1 of \citet{Ab18} is 0.3 per cent of the flux of the Crab Nebula, which was measured by H.E.S.S. to be F($>$1 TeV) $= 2.26 \times 10^{-11}$ cm$^{-2}$ s$^{-1}$  \citep{Ah06} and gives the upper limit on the TeV $\gamma$-ray efficiencies $\eta_{\rm TeV} < 10^{-3.6}$.

The obtained efficiencies can be compared with those of the other young PWNe. Based on the spectral modeling by \citet{TaTa10,TaTa11,TaTa13}, the efficiencies range as $10^{-3.9} \la \eta_{\rm TT, keV} \la 10^{-1.7}$, $10^{-4.7} \la \eta_{\rm TT, GeV} \la 10^{-3.7}$, and $10^{-5.0} \la \eta_{\rm TT, TeV} \la 10^{-3.4}$. The PWN has a typical X-ray efficiency and the upper limit on the TeV $\gamma$-ray efficiency is not strong enough to rule out the leptonic emission from the PWN. The observed GeV efficiency is fairly high compared with the leptonic emission model of young PWNe. In addition, the lower limit on the GeV to TeV flux ratio $\eta_{\rm GeV} / \eta_{\rm TeV} \ga 10^{2.2}$ is also much higher than that of the GeV PWNe candidates $\la 10$ (Fig. 12 of \citealp{Ac13}). We conclude that the observed GeV photons do not likely come from the PWN.

Assuming the distance of 6.2 kpc, the 0.2$-$300 GeV luminosity of $1.4\times10^{35}$ erg s$^{-1}$ is close to the bright end of the SNRs observed by {\it Fermi} (Figs. 17 and 18 of \citealp{Ac16}). Young SNRs are often dim in the 1$-$100 GeV energy band. Their faint GeV emission can be best explained by the IC scattering off the CMB photons by relativistic electrons \citep{Ac16}. The high ($\ga 10^{35}$ erg s$^{-1}$) GeV luminosity of 3C 396 is much more difficult to be reproduced by this mechanism. However, in the energy range of 1$-$100 GeV it is consistent with the high luminosity values found in old SNRs interacting with MCs which emit gamma rays likely through the hadronic mechanism. 

The bremsstrahlung process may be effective in places where the density of the ambient medium is high ($n$ $>$ 100 cm$^{-3}$). In addition, hadronic interactions could dominate over the bremsstrahlung process, if the ratio of energy densities of protons to electrons ($W_{\rm p}/W_{\rm e}$) is greater than 10, which corresponds to the case when the gamma-ray emitting region is located far from the acceleration region \citep{Ta18}. Therefore, in order to reveal the dominating gamma-ray emission mechanism, multi-wavelength data, e.g. radio data, need to be investigated together with the GeV results we present in this paper. 

%%%CONCLUSION
\section{Conclusions}
\label{conclusions}
We have examined the X-ray emission from the SNR 3C 396 by studying its spectra with {\it Suzaku}. We also investigated GeV gamma-ray emission from the SNR as observed by {\it Fermi}-LAT. The conclusions of our work can be described as follows.

\begin{easylist}[itemize]
\vspace{7pt}
&  Using {\it Suzaku} data, we found that the shell spectrum is clearly thermal ($\sim$1.12 keV), with no sign of a non-thermal component. The slightly enhanced abundances of Al and Ca found in the SNR shell suggest that the X-ray plasma is likely to be ejecta origin. The PWN emission is characterized by a power-law model with a photon index of $\sim$1.97 and thermal emission with an electron temperature of $\sim$0.93 keV.

& In the energy range of 1$-$300 GeV, we detected an excess of gamma rays associated with 3C 396 / 4FGL J1903.8+0531 and calculated the best-fit position to be R.A.(J2000) = 285$^{\circ}\!\!$.957 $\pm$ 0$^{\circ}\!\!$.027 and decl.(J2000) = 5$^{\circ}\!\!$.475 $\pm$ 0$^{\circ}\!\!$.021. 

& In the energy range of 0.2$-$300 GeV, 3C 396 / 4FGL J1903.8+0531 was detected with a significance of 18$\sigma$ at the best-fit position, where it was fit with a log-parabola-type spectrum having indices of $\alpha$=2.66 and $\beta$=0.16. 

& If gamma rays are produced through the leptonic PWN scenario, TeV gamma-ray emission should be observed from 3C 396, which have not been detected so far. In addition, assuming a distance of 6.2 kpc, the luminosity of 3C 396 / 4FGL J1903.8+0531 was found to be too high ($\ga 10^{35}$ erg s$^{-1}$) to be explained by the IC emission model. So, alternative gamma-ray emission mechanisms may be the bremsstrahlung process and the hadronic model. The location and the spectral features of 3C 396 / 4FGL J1903.8+0531 indicate that this SNR might be expanding into a dense MC at northern and western sides of the SNR's shell and accelerated protons might be penetrating these dense clouds creating hadronic gamma rays. However, the leptonic contribution (i.e. bremsstrahlung emission) to the total gamma-ray emission could be significant. 

& In our second paper, we will concentrate on spectral modelling of 3C 396 / 4FGL J1903.8+0531, especially using radio data, to find out if the dominating gamma-ray emission mechanism in 3C 396 is leptonic (i.e. bremsstrahlung) or hadronic in origin.

\end{easylist}

%%% ACKNOWLEDGEMENTS
\section*{Acknowledgments}
We would like to thank Dr. Aya Bamba, and all the {\it Suzaku} team members. We also thank the anonymous referee for useful comments and suggestions. AS is supported by the Scientific and Technological Research Council of Turkey (T\"{U}B\.{I}TAK) through the B\.{I}DEB-2219 fellowship program. This work is supported in part by grant-in-aid from the Ministry of Education, Culture, Sports, Science, and Technology (MEXT) of Japan, No.18H01232(RY), No.16K17702(YO), No.17K18270(ST) and 16J06773, 18H01246 (SK). We are also grateful to the {\it Fermi} team for making the {\it Fermi}-LAT data and analysis tools available to the public.

$~$

%%% FACILITIES
{\it Facility}: {\it Suzaku} and {\it Fermi}-LAT. 
%\newpage

%%% BIBLIOGRAPHY

\onecolumn

\twocolumn

%%% END OF DOCUMENT
\end{document}